\documentclass[a4paper,11pt]{article}
\usepackage{indentfirst}
\usepackage{mathtools}
\usepackage{verbatim}
\usepackage{float}
\usepackage{booktabs}

\usepackage{bm}
\usepackage{amsmath}
\usepackage{amssymb}

\usepackage{subcaption}
\usepackage{multirow}
\usepackage{longtable}
\allowdisplaybreaks[1]

\usepackage[utf8]{inputenc}
\usepackage[T1]{fontenc}

\usepackage{graphicx}

\usepackage{hyperref}
\usepackage{xcolor}

\hypersetup{
    colorlinks,
    linkcolor={red!50!black},
    citecolor={blue!50!black},
    urlcolor={blue!80!black}
}

\usepackage{comment}

\usepackage{braket}

\usepackage{geometry}
\usepackage{cancel}
\usepackage{physics}

\usepackage[nottoc,notlot,notlof]{tocbibind}	

\usepackage[toc,page]{appendix}     
\usepackage[inline]{enumitem}

\usepackage{tikz}

\usepackage{authblk}

\begin{document}

\title{Volume Changing Symmetries by Matrix Product Operators}
\author[1]{M\'arton Borsi}
\author[1]{Bal\'azs Pozsgay}
\affil[1]{MTA-ELTE “Momentum” Integrable Quantum Dynamics Research Group, \protect\\
  ELTE E\"otv\"os Lor\'and University, \protect\\ Budapest, Hungary}

\date{}

\maketitle

\begin{abstract}
We consider spin chain models with exotic symmetries that change the length of the spin chain. It is known that the
XXZ Heisenberg spin chain at the supersymmetric point $\Delta=-1/2$ possesses such a symmetry: it is given by the
supersymmetry generators, which change the length of the chain by one unit. We show that volume changing symmetries
exist also in other spin chain models, and that they can be constructed using a special tensor network, which is a
simple generalization of a  Matrix Product Operator.
As examples we consider the folded XXZ model and its perturbations, and also a new hopping model that is defined on
constrained Hilbert spaces. We show that the volume changing symmetries are not related to integrability: the symmetries
can survive even non-integrable perturbations. We also show that the known supersymmetry generator of the XXZ chain with 
$\Delta=-1/2$ can also be expressed as a generalized Matrix Product Operator.
\end{abstract}

\section{Introduction}

Symmetries play a crucial role in many branches of physics. According to Noether's theorem, continuous symmetries are
associated with conservation laws, which affect the dynamical processes in many body systems.
It is a central question
in theoretical physics to determine what types of symmetries can exist in different types of models.

In quantum mechanics symmetries are represented by operators that commute with a Hamiltonian. If a model has additional
symmetries on top of the usual ones, then this leads to unconventional dynamical behaviour.
Extra conservation laws imply the breaking of ergodicity, often leading to exotic dynamical signatures.

It has been a central topic in the last two decades to identify and analyze various ways of ergodicity breaking and the
types of symmetries that can occur in quantum many body systems.
Unconventional symmetries appear in the so-called integrable models, which possess an infinite set
of commuting conserved quantities, which are extensive with local densities 
\cite{sutherland-book,Korepin-Book}.
Many-body localization is another mechanism, where
strong disorder leads to the emergence of an extensive set of local integrals of motion \cite{huse-review}.
Exotic symmetries are seen also in  models with Hilbert space fragmentation
\cite{hfrag-review,fragmentation-scars-review-2};
such models have a symmetry algebra whose dimension also grows exponentially with the volume
\cite{fragm-commutant-1}.

In one spatial dimension the exotic symmetries can often be represented by a Matrix Product Operator (MPO). An MPO is a
one dimensional tensor network, constructed out of fourth rank tensors. The operator space entanglement of an MPO is
bounded by 
the bond dimension, which is a fixed number of the network. Therefore, such MPO's can be seen as a ``sufficiently local''
generalization of the conventional symmetries, which are described by symmetry operations that are stricly local.

A new type of MPO symmetry was found recently in  \cite{sajat-mpo}, and it was demonstrated that this symmetry also
leads to the breaking of ergodicity: the models display unusual dynamical properties, such as the presence of persistent
oscillations. It was also shown in   \cite{sajat-mpo} that these MPO symmetries are not related to integrability: they
can appear in integrable and in non-integrable models as well. They have their origin in certain hidden ``geometric''
properties of the Hamiltonian. 

There is one family of unconventional symmetries which has received relatively little attention: those symmetries of
spin chains that 
change the length of a chain. Such symmetries connect Hilbert spaces of spin chains with different lengths. In these
cases there are separate Hamiltonians acting on Hilbert spaces with different length, but the Hamiltonians have the same
operator density. Volume changing symmetries can then be seen as intertwiners for the different Hamiltonians, and the
symmetry  leads to spectral degeneracies across different volumes.

Up to our best knowledge,
there is only one example in the literature for such a symmetry, and this is a certain type of supersymmetry discovered
in \cite{susy-yang-fendley}. Even earlier it was known that lattice models can possess supersymmetry
\cite{susy-constrained-lattice-hopping1,susy-constrained-lattice-hopping2},
 but in these earlier models the supersymmetry operation acts within the Hilbert space of the
model. It was understood later in \cite{susy-yang-fendley} that the XXZ Heisenberg spin chain also possesses
supersymmetry at the special point $\Delta=-1/2$. It was shown, that in this case the supersymmetry generators do change
the length of the spin chain, and an explicit real space representation was given for the generators. A representation
using Algebraic Bethe Ansatz was given later in \cite{weston-susy-aba}. In \cite{xyz-susy-1} it was also shown that
volume changing supersymmetry operators exist also in the XYZ chain, if the couplings obey a special relation. For
further generalizations see \cite{hagendorf-dynamical-susy,higher-rank-lattice-susy,hagendorf-dynamical-susy-2}.

In this work we revisit the models studied in \cite{sajat-mpo} and show that they also possess volume changing
symmetries, and that these operators are described by tensor networks, that are very simple generalizations of MPO's. In
fact, they can be seen as an MPO with an additional boundary tensor that adds one more site to the given chain. The list
of models with such symmetries includes the so-called folded XXZ model \cite{folded0,folded1,folded2,sajat-folded} (see
also \cite{folded-XXZ-stochastic-1}), and multiple types of perturbations of this model, such as the so-called hard rod
deformed XXZ models  \cite{sajat-hardrod} and a large family of non-integrable perturbations as well. Furthermore, in
this work we present a new family of models with such symmetries; these are models defined on constrained Hilbert spaces.

We should note that volume changing operators appear in the literature dealing with the AdS/CFT duality
\cite{beisert20082}. The spin chains that describe scaling dimensions of composite operators in the planar limit of
$\mathcal{N}=4$ SCFT's are ``dynamical'': they connect Hilbert spaces of spin chains with different length. Up to our
best knowledge, the integrability of these Hamiltonians is not yet understood, and any progress made on volume changing
symmetries could lead to a better understanding of the integrable structure of these Hamiltonians.

The structure of this paper is as follows. In Section \ref{sec:symdef} we introduce the volume changing symmetries and
the tensor networks that will be used to construct them. In Section \ref{sec:models} we introduce the models we study,
namely the folded XXZ model and a new 5-site interacting hopping model.
In Section \ref{sec:MM} we show how to map them to the Maassarani-Mathieu (MM) model via a non-local
transformation. This 
mapping was already published in multiple works (see for example \cite{sajat-folded,sajat-mpo}), and it provides the
basis for studying the MPO symmetries. In \ref{sec:MMsym} we revisit the MPO symmetries of the MM model. In
Sections \ref{sec:folded} and \ref{sec:5site} we construct volume changing symmetries for our two selected
models. In Section \ref{sec:nonint} we explain that the MPO symmetries are kept intact by selected non-integrable
perturbations. In Section \ref{sec:susy} we show that the volume changing lattice supersymmetry first derived in
\cite{susy-yang-fendley} can also be written as an MPO using our techniques. Finally, we present our Conclusions in
Section \ref{sec:concl}.

\section{Volume Changing Symmetries}

\label{sec:symdef}

Our goal is to construct symmetry operators that alter the volume of the system. We will focus on one dimensional spin
chain models. In all cases we consider Hamiltonians with open boundary conditions defined as
\begin{equation}
  H_L=\sum_{j} h(j),
\end{equation}
where $h(j)$ is an operator density and $L$ denotes the length of the chain. The limits on the summation depend on the
details of the model (such as the range 
of the Hamiltonian density).  We will focus on spin-1/2 chains, but
examples with higher local dimensions could be constructed with similar ideas. We do not treat periodic boundary
conditions. 

Let us denote by $\mathcal{H}_L$ the Hilbert space of the model at length $L$. We will treat models that have a volume
changing symmetry $\mathcal{S}_L$. In our examples the symmetry operation adds one site to the chain, therefore it will be
described by a linear operator that acts from $\mathcal{H}_L$ to $\mathcal{H}_{L+1}$. It is a symmetry in the following
generalized sense. First of all, the symmetry intertwines Hamiltonians defined in different volumes:
\begin{equation}
    \mathcal{S}_L H_L = H_{L+1} \mathcal{S}_L.
\end{equation}
  
Alternatively, we can also consider the direct sum of Hilbert spaces $\mathcal{H}_L$ corresponding to different volumes
$L$ of the given model: 
\begin{equation}
    \mathcal{H} = 
   \bigoplus_{L=L_{\text{min}}}^\infty \mathcal{H}_L,
\end{equation}
where $L_{\text{min}}$ is the number of sites the Hamiltonian density acts on so that the model can actually exist in such a volume.

The combined Hamiltonian $H$ defined on this enlarged Hilbert space is also given as the direct sum of Hamiltonians $H_L$ corresponding to different volumes:
\begin{equation}
    H = 
   \bigoplus_{L=L_{\text{min}}}^\infty H_L,
\end{equation}
so that it is block diagonal and does not connect different sectors.

The combined symmetry operators $\mathcal{S}= \bigoplus_{L=L_{\text{min}}}^\infty \mathcal{S}_L$ act off-diagonally
between the sectors, and they commute with the combined 
Hamiltonian: 
\begin{equation}
    [ \mathcal{S}, H ] = 0
\end{equation}

\subsection{Matrix Product Operators}

We will consider symmetry operators that are built as Matrix Product Operators (MPO). First we consider the standard case,
where the MPO acts on a chosen Hilbert space. 

\begin{figure}
     \centering
     \begin{subfigure}[b]{0.3\textwidth}
         \centering
         \includegraphics[height=1.1 cm]{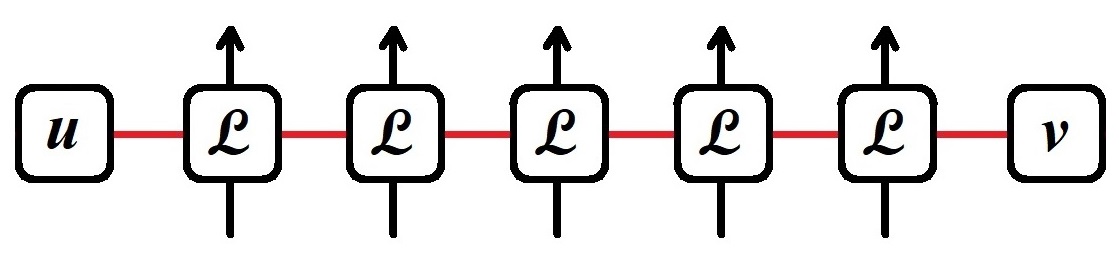}
         \caption{}
         \label{fig:open-construction1}
     \end{subfigure}
     \hfill
     \begin{subfigure}[b]{0.3\textwidth}
         \centering
         \includegraphics[height=1.1 cm]{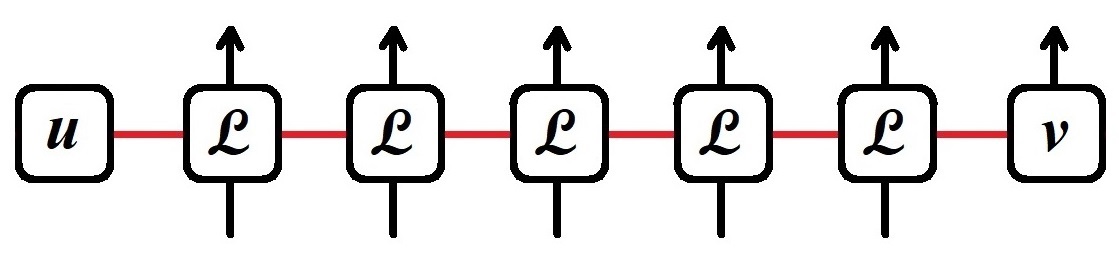}
         \caption{}
         \label{fig:open-construction2}
     \end{subfigure}
     \hfill
     \begin{subfigure}[b]{0.3\textwidth}
         \centering
         \includegraphics[height=1.1 cm]{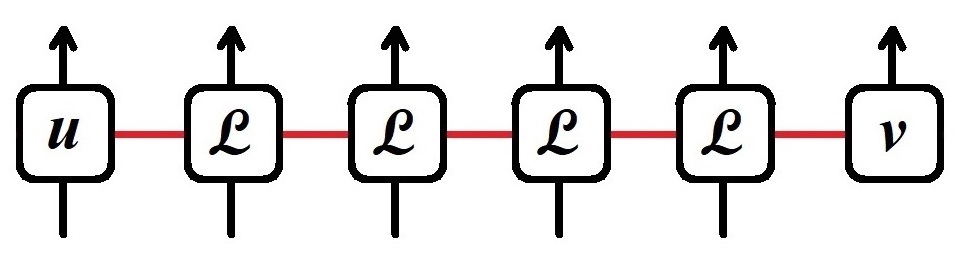}
         \caption{}
         \label{fig:open-construction3}
     \end{subfigure}
        \caption{MPO constructions for open boundary conditions. A volume preserving operator is produced by (a), whereas (b) and (c) are two alternative ways to add an extra site to the system.}
        \label{fig:open-construction}
\end{figure}

Let us introduce the elementary operator $\mathcal{L}_{a,j}$ that acts on the tensor product
space $V_a \otimes V_j$ with $V_a$ being the auxiliary space and $V_j$ the physical one corresponding to site
$j$. This can also be interpreted as a four-leg (fourth rank) tensor. The MPO with open boundary conditions also requires 
 boundary vectors $u,v \in V_a$. Then the MPO is given by 
\begin{equation}
  \label{MPO}
    \mathcal{S} = v^T \mathcal{L}_{a,L} \mathcal{L}_{a,L-1}...\ \mathcal{L}_{a,1} u,
\end{equation}
where $v^T$ is the transpose. The resulting operator $\mathcal{S}$  only acts on the physical space. The usual graphical notation for this construction is shown in Figure \ref{fig:open-construction1}. Horizontal legs correspond to the auxiliary space whereas vertical legs to the physical ones.

The simple volume changing effect of adding a single site to the system can be achieved by a slight modification of the construction above. The formula \eqref{MPO} remains intact, but the boundary vector $v$ is now an element of $V_a \otimes V_{L+1}$ so that it produces the extra site. This is shown in Figure \ref{fig:open-construction2}.

We also introduce an alternative construction where the action on the first site is not performed by $\mathcal{L}_{a,1}$
but the boundary vector $u$. Formally $u$ is an element of $V_a\times \text{End}(V_{1})$, and the MPO is built as the
product 
\begin{equation}
    \mathcal{S} = v^T \mathcal{L}_{a,L} \mathcal{L}_{a,L-1}...\ \mathcal{L}_{a,2} u
    \label{MPO2}
\end{equation}
that is presented in Figure \ref{fig:open-construction3} using the graphical notation. This construction can produce the same MPOs as the previous one but it provides more convenient formulas in certain cases.

\section{Models}

\label{sec:models}

Here we introduce the two models for which the volume changing symmetry operators will be constructed. Both are
integrable spin-1/2 lattice models primarily considered with open boundary conditions.
Beside the usual spin operators we use the local projectors $P_j=(1+S_j^z)/2$ and $N_j=(1-S_j^z)/2$. 

The first model is the folded XXZ chain defined by the 4-site Hamiltonian density
\begin{equation}
    h_{\text{fXXZ},j} = P_j(S_{j+1}^+ S_{j+2}^- + S_{j+1}^- S_{j+2}^+) P_{j+3} + N_j(S_{j+1}^+ S_{j+2}^- + S_{j+1}^- S_{j+2}^+) N_{j+3}.
\end{equation}
This Hamiltonian density generates the following transitions between segments on four sites:
\begin{equation}
    \ket{\uparrow\downarrow\uparrow\uparrow} \leftrightarrow \ket{\uparrow\uparrow\downarrow\uparrow} \qquad \text{and} \qquad \ket{\downarrow\uparrow\downarrow\downarrow} \leftrightarrow \ket{\downarrow\downarrow\uparrow\downarrow}.
\end{equation}
This means, that  a single up or down spin is able to propagate in the sea of opposing spins. The model is invariant to
global spin flips and the total magnetization is conserved. 

Our second model is defined on a constrained Hilbert space: From a standard tensor product space we allow only those
basis states in the computational basis, which do not have two neighbouring down spins. This is often called the Rydberg
constraint, because it naturally arises in experiments with Rydberg atoms \cite{rydberg-blockade-experimentally}. 
A  minor technical point: we also exclude states that have down spins at the boundary (first and last sites).
As an example we present a representative of the allowed states:
\begin{equation}
    \ket{\uparrow\downarrow\uparrow\uparrow\downarrow\uparrow\uparrow\uparrow\uparrow\downarrow\uparrow\uparrow\uparrow\downarrow\uparrow\uparrow}.
\end{equation}

Finally, our second model is described by the 5-site Hamiltonian density
\begin{equation}
    h_{\text{5-site},j} = P_j (S_{j+1}^+ P_{j+2} S_{j+3}^- + S_{j+1}^- P_{j+2} S_{j+3}^+) P_{j+4} + N_j (S_{j+1}^+ N_{j+2} S_{j+3}^- + S_{j+1}^- N_{j+2} S_{j+3}^+) N_{j+4}.
\end{equation}
This operator density generates the local exchanges
\begin{equation}
    \ket{\uparrow\downarrow\uparrow\uparrow\uparrow} \leftrightarrow \ket{\uparrow\uparrow\uparrow\downarrow\uparrow} \qquad \text{and} \qquad \ket{\downarrow\uparrow\downarrow\downarrow\downarrow} \leftrightarrow \ket{\downarrow\downarrow\downarrow\uparrow\downarrow}.
\end{equation}
The extra control bit added in the middle ensures that single up and down spins can only propagate on the even/odd
sublattices. As a result the total magnetization on the two sublattices are conserved individually. It is clear that the
Rydberg constraint is also conserved.

This model first appeared recently in the paper \cite{sajat-rydberg}, which performed a partial classification of
integrable models on Rydberg constrained Hilbert spaces\footnote{The Hamiltonian appeared below eq. (186) of
  \cite{sajat-rydberg}, in Appendix A.}. It can be shown using the methods of
\cite{sajat-medium,sajat-rydberg} that the model is integrable.
This claim is not relevant to our discussion, therefore we do not present a detailed proof here. 

In the following Sections we construct volume changing symmetries for these models and their perturbations.
Our strategy is the same as in the previous work \cite{sajat-mpo}. First we show that both spin chains can
be mapped to the Maassarani-Mathieu (MM) model using a non-local mapping. The symmetries of the MM  model are easy to
understand and we build the 
symmetry generators  as matrix product operators (MPOs). Finally we build the MPO symmetry
operators in the original models by retracing the mappings. We will explain that non-local nature of the mapping to the
MM model (and the inverse of this mapping) introduces the volume changing effect.

\section{Mapping to the Maassarani-Mathieu Model}

\label{sec:MM}

First we perform a bond-site transformation on the two spin models. The idea is to place variables on the bonds between
sites according to the following rule. We consider the computational basis, and for each bond we write down a local 
state $\ket{\circ}$ or $\ket{\bullet}$, if the neighbouring spins are identical or opposite, respectively.
Both Hamiltonians are invariant to spin reflection thus this mapping is well
defined for the models. For more details see  \cite{sajat-folded,sajat-mpo}.

The folded XXZ chain has the bond Hamiltonian density
\begin{equation}
    h_{\text{fXXZ},j}^\text{bond} = \sigma_j^+ P_{j+1}^\bullet \sigma_{j+2}^- + \sigma_j^- P_{j+1}^\bullet \sigma_{j+2}^+,
\end{equation}
where we used the local projectors $P^\circ=\ketbra{\circ}{\circ}$ and $P^\bullet=\ketbra{\bullet}{\bullet}$. The
transitions generated by this Hamiltonian density are
\begin{equation}
    \ket{\bullet\bullet\circ} \leftrightarrow \ket{\circ\bullet\bullet}.
    \label{eq:bond-rule1}
\end{equation}
Similarly the 5-site Hamiltonian density becomes
\begin{equation}
    h_{\text{5-site},j}^\text{bond} = \sigma_{j}^+ \sigma_{j+1}^+ \sigma_{j+2}^-\sigma_{j+3}^- +\sigma_{j}^- \sigma_{j+1}^- \sigma_{j+2}^+\sigma_{j+3}^+
\end{equation}
corresponding to the exchange
\begin{equation}
    \ket{\bullet\bullet\circ\circ} \leftrightarrow \ket{\circ\circ\bullet\bullet}.
    \label{eq:bond-rule2}
\end{equation}
The original constraint that we have single down spins in the sea of up spins results in the states $\ket{\bullet}$ always coming in pairs, that is, only an even number of successive $\ket{\bullet}$ states is allowed.

Our goal is to map both of these bond Hamiltonians to the Maassarani-Mathieu (MM) chain. The MM model is a spin chain
model with 
local dimension 
three \cite{su3-xx,XXC}. The state $\ket{0}$ is considered to be the vacuum whereas $\ket{1}$ and $\ket{2}$ are regarded as particles with an internal degree of freedom. The Hamiltonian describes the propagation of these particles in the vacuum allowing the transitions
\begin{equation}
    \ket{01} \leftrightarrow \ket{10} \qquad \text{and} \qquad \ket{02} \leftrightarrow \ket{20}.
\end{equation}
The corresponding Hamiltonian density is
\begin{equation}
    h_{\text{MM},j} = \sum_{a=1}^2 ( E_j^{0,a} E_{j+1}^{a,0} + E_j^{a,0} E_{j+1}^{0,a} ),
\end{equation}
where $E_{j}^{a,b}$, with $a,b \in \{ 0,1,2 \}$, are matrices acting at site $j$ that contain a single $1$ at position $(a,b)$.

We map the bond models to the MM chain by identifying local segments of the bond models with the basis states 
$\ket{0}$, $\ket{1}$ and $\ket{2}$. The local segments will have varying length, and this is a key non-local property of
the mapping.
There are two conditions for the mapping to work:
there should be a one-to-one correspondence between each bond state and the MM states, and
the Hamiltonians should be mapped to each other.
 
For the folded XXZ chain such an identification is given by \cite{sajat-folded,sajat-mpo}
\begin{align}
    \ket{0} &= \ket{\bullet\bullet} \\
    \ket{1} &= \ket{\circ} \\
    \ket{2} &= \ket{\bullet\circ}.
    \label{eq:identification1}
\end{align}
Note that particles always end in $\ket{\circ}$ therefore the exchange rule \eqref{eq:bond-rule1} indeed corresponds to a particle-vacuum exchange.

In the case of the 5-site model we use the identification
\begin{align}
    \ket{0} &= \ket{\circ\circ} \\
    \ket{1} &= \ket{\bullet\bullet} \\
    \ket{2} &= \ket{\circ\bullet\bullet}.
    \label{eq:identification2}
\end{align}
Particles now end in $\ket{\bullet\bullet}$ thus, according to the exchange rule \eqref{eq:bond-rule2},
the correct Hamiltonian is obtained again.

We remark that there is a connection between our two models. If
we exchange the two local states $\ket{\circ} \leftrightarrow \ket{\bullet}$ for the 5-site model in the bond picture,
then the states $\ket{0}$ 
become identical and the only difference is that particles $\ket{1}$ and $\ket{2}$ are elongated, ending with
$\ket{\circ\circ}$ instead of $\ket{\circ}$.

\section{Symmetries in the MM Model}

\label{sec:MMsym}

In the MM model we can regard the local states $\ket{1}$ and $\ket{2}$  as particles with up and down spin,
respectively. Examining the Hamiltonian of the model we find that the spin of a particle
does not influence the 
dynamics in any way, and the sequence of the successive spin values does not change throughout the time evolution, since
particles are unable to overtake each other. This is an example of spin-charge separation with the positions of the
particles determining the charge part.  The spin part of the wave function is a constant of motion, and  
in the computational basis this is also called the irreducible string \cite{folded-XXZ-stochastic-1}.
Different values of the irreducible string label dynamically disconnected
sectors in the Hilbert space, and this results in the phenomenon of Hilbert space fragmentation
\cite{sajat-folded,sajat-mpo}. The spin-charge separation of the MM model and in certain closely related cellular
automata leads to anomalous diffusion \cite{prosen-anomalous-universal}.

A consequence of the statements above is that any operator that only modifies the irreducible string, that is, performs
exchanges of the form $\ket{1} \leftrightarrow \ket{2}$, is a symmetry of the Hamiltonian. In principle one can
construct these symmetry operators in the original spin models as well by following the mappings in the inverse
direction. However, the effect of the non-local property can not be ignored. 

The states $\ket{1}$ and $\ket{2}$ correspond to local bond states that span different number of sites thus performing
exchanges of these two introduces a change in the volume in general. In our previous work \cite{sajat-mpo} we avoided
this problem by considering operators which always implement the exchanges $\ket{1}\rightarrow\ket{2}$ and
$\ket{2}\rightarrow\ket{1}$ in pairs. In this paper we present the construction of volume changing operators through the
simple example of a single $\ket{1}\rightarrow\ket{2}$ exchange which adds one extra site in the bond and spin
models. This forces us to use open boundary conditions.

\begin{figure}[h]
    \centering
    \includegraphics[width=10 cm]{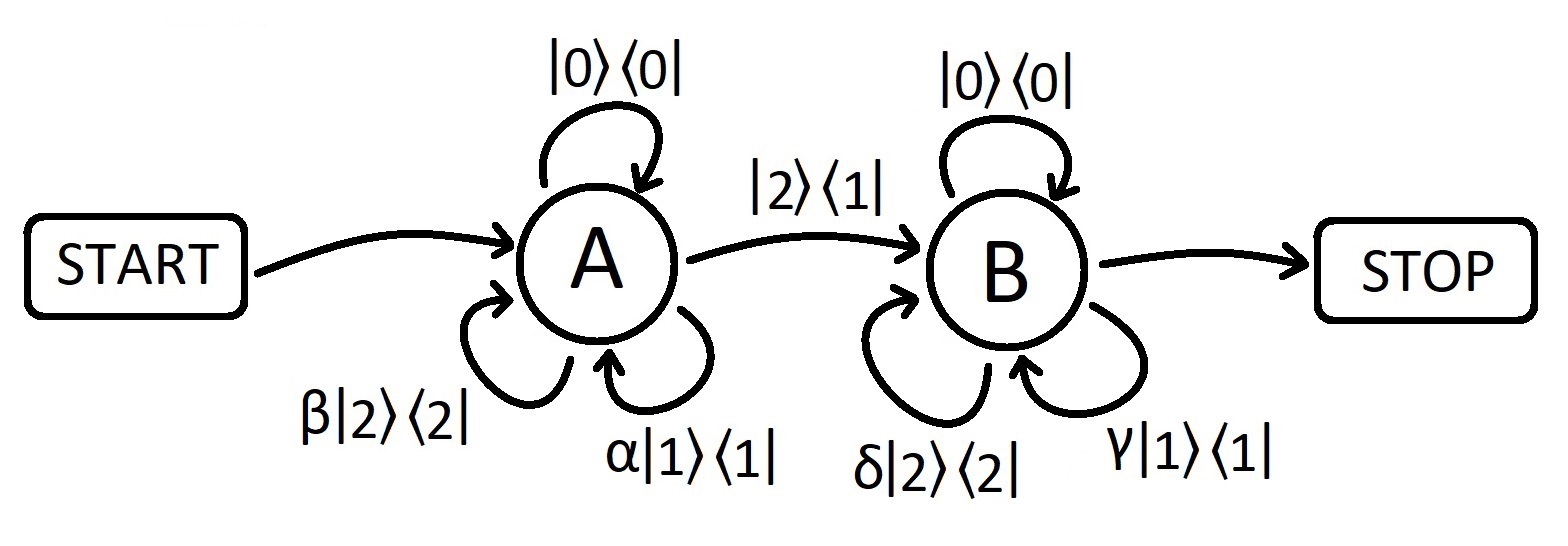}
    \caption{Automaton for the MM model.}
    \label{fig:mpo-mm}
\end{figure}

 We use the automaton picture of MPOs \cite{mps-automata} where each state of the automaton correspond to an auxiliary
 dimension and each transition between these indicates the corresponding entry of the elementary matrix. Such an
 automaton offers a graphical interpretation about the action of the operator. A simple example is given by Figure
 \ref{fig:mpo-mm}. describing the MPO performing the single $\ket{1}\rightarrow\ket{2}$ exchange. 

The process is explained as follows. Automaton state $A$ performs diagonal action until the exchange $\ket{1}\rightarrow\ket{2}$ happens followed by repeated diagonal action by automaton state $B$. The elementary tensor and the boundary vectors of the MPO \eqref{MPO} can be read off:
\begin{align}
    \mathcal{L}_\text{MM} &= \ketbra{0}{0}
    \begin{bmatrix}
        1 & 0 \\
        0 & 1 \\
    \end{bmatrix}
    + \ketbra{1}{1}
    \begin{bmatrix}
        \alpha & 0 \\
        0 & \gamma \\
    \end{bmatrix}
    + \ketbra{2}{2}
    \begin{bmatrix}
        \beta & 0 \\
        0 & \delta \\
    \end{bmatrix}
    + \ketbra{2}{1}
    \begin{bmatrix}
        0 & 0 \\
        1 & 0 \\
    \end{bmatrix}
    \\
    u_\text{MM} &=
    \begin{bmatrix}
        1 & 0 \\
    \end{bmatrix}
    \\
    v_\text{MM} &=
    \begin{bmatrix}
        0 & 1 \\
    \end{bmatrix}.
\end{align}
For now, the construction \ref{fig:open-construction1} is used and the volume is preserved. All matrix and vector notations correspond to the auxiliary space whereas the physical space action is indicated by the bra-ket notation. Greek letters denote free complex nonzero parameters resulting in a whole family of symmetry operators. They indicate the following:
\begin{align}
    \alpha:& \text{ particle $\ket{1}$ before the exchange} \\
    \beta:& \text{ particle $\ket{2}$ before the exchange}\\
    \gamma:& \text{ particle $\ket{1}$ after the exchange}\\
    \delta:& \text{ particle $\ket{2}$ after the exchange}\\
\end{align}
The coefficients of the $\ketbra{0}{0}$ operators must be identical in order to have commutation with the Hamiltonian, and they can be set to one. Coefficients corresponding to the exchange $\ketbra{2}{1}$ and to the boundary vectors can also be set to one by overall normalization.

Our goal now is to construct MPOs that have analogous action in the original models. We achieve this in two steps:
retracing the mapping we first build the symmetry operators for the bond models then for the spin chains.

\section{Symmetries in the Folded XXZ Model}

\label{sec:folded}

\begin{figure}[h]
    \centering
    \includegraphics[width=12 cm]{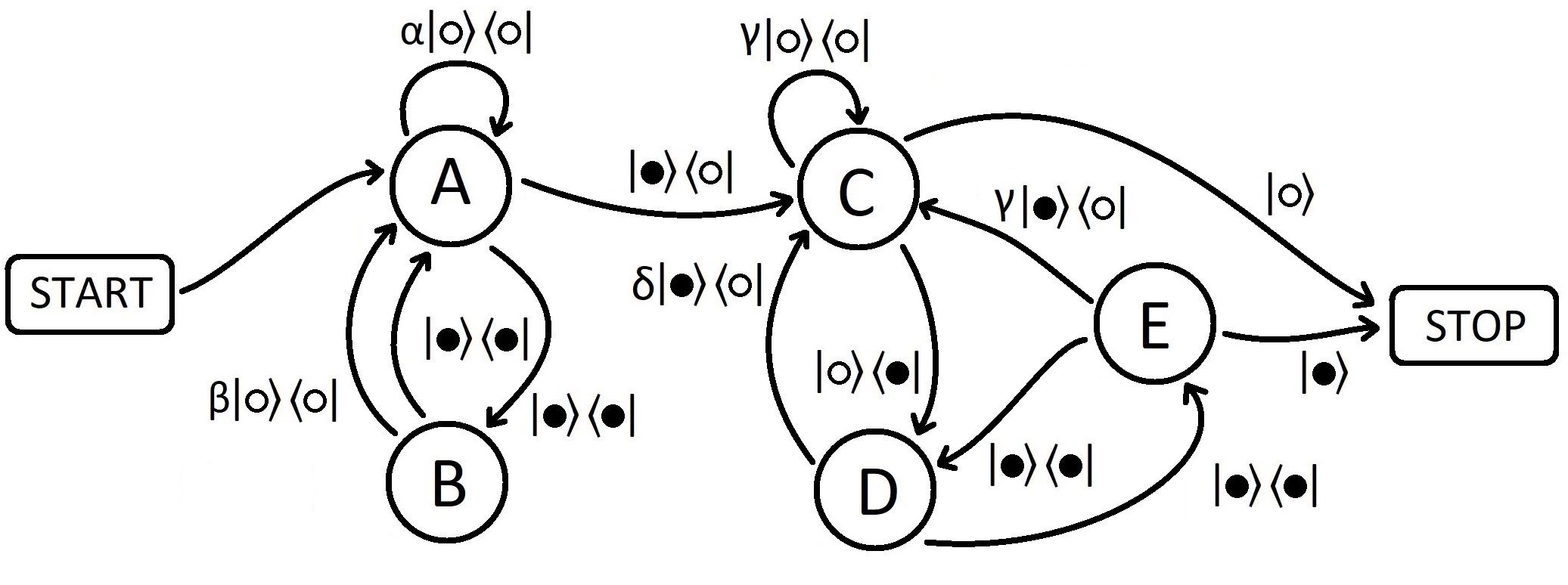}
    \caption{Automaton for the folded XXZ bond model.}
    \label{fig:fxxz-bond}
\end{figure}

The basic idea is to build an automaton with the same units as in the MM model: diagonal action first, then performing the exchange and continue with diagonal actions again. The difference is that MM local states might span multiple sites in the bond model. The initial diagonal action stays relatively simple but upon the exchange we need to replace $\ket{1}=\ket{\circ}$ with $\ket{2}=\ket{\bullet\circ}$. The latter spans one extra site, thus, instead of the simple diagonal action, each following site is to be shifted to the right. At the very end we add one extra site to the lattice in order to perform the final shift. The corresponding automaton is shown in Figure \ref{fig:fxxz-bond}.

Automaton states $A$ and $B$ are responsible for the initial diagonal action which is followed by the transition from $A$ to $C$. This performs the exchange followed by the one-site shift for all the states to the right managed by automaton states $C$, $D$ and $E$. Note that throughout the process the identification of the three MM states is indicated by the various constants. In the last step we add the extra site with the correct state.

Reading off the matrix and vector elements we arrive at
\begin{align}
\begin{split}
    \mathcal{L}_\text{fXXZ}^\text{bond} =& \ketbra{\circ}{\circ} 
    \begin{bmatrix}
        \alpha & \beta & 0 & 0 & 0 \\
        0 & 0 & 0 & 0 & 0 \\
        0 & 0 & \gamma & 0 & 0 \\
        0 & 0 & 0 & 0 & 0 \\
        0 & 0 & 0 & 0 & 0 \\
    \end{bmatrix}
    + \ketbra{\bullet}{\bullet}
    \begin{bmatrix}
        0 & 1 & 0 & 0 & 0 \\
        1 & 0 & 0 & 0 & 0 \\
        0 & 0 & 0 & 0 & 0 \\
        0 & 0 & 0 & 0 & 1 \\
        0 & 0 & 0 & 1 & 0 \\
    \end{bmatrix}
    \\
    + & \ketbra{\bullet}{\circ}
    \begin{bmatrix}
        0 & 0 & 0 & 0 & 0 \\
        0 & 0 & 0 & 0 & 0 \\
        1 & 0 & 0 & \delta & \gamma \\
        0 & 0 & 0 & 0 & 0 \\
        0 & 0 & 0 & 0 & 0 \\
    \end{bmatrix}
    + \ketbra{\circ}{\bullet}
    \begin{bmatrix}
        0 & 0 & 0 & 0 & 0 \\
        0 & 0 & 0 & 0 & 0 \\
        0 & 0 & 0 & 0 & 0 \\
        0 & 0 & 1 & 0 & 0 \\
        0 & 0 & 0 & 0 & 0 \\
    \end{bmatrix}
\end{split}
    \\
    u_\text{fXXZ}^\text{bond} = &
    \begin{bmatrix}
        1 & 0 & 0 & 0 & 0 \\
    \end{bmatrix}
    \\
    v_\text{fXXZ}^\text{bond} = &
    \begin{bmatrix}
        0 & 0 & \ket{\circ} & 0 & \ket{\bullet} \\
    \end{bmatrix}.
\end{align}
The MPO is constructed according to Figure \ref{fig:open-construction2} and formula \eqref{MPO}. We remark that this MPO annihilates the aforementioned bond states that can not be interpreted as MM states due to the odd number of $\ket{\bullet}$'s on the right.

\begin{figure}[h]
    \centering
    \includegraphics[width=12 cm]{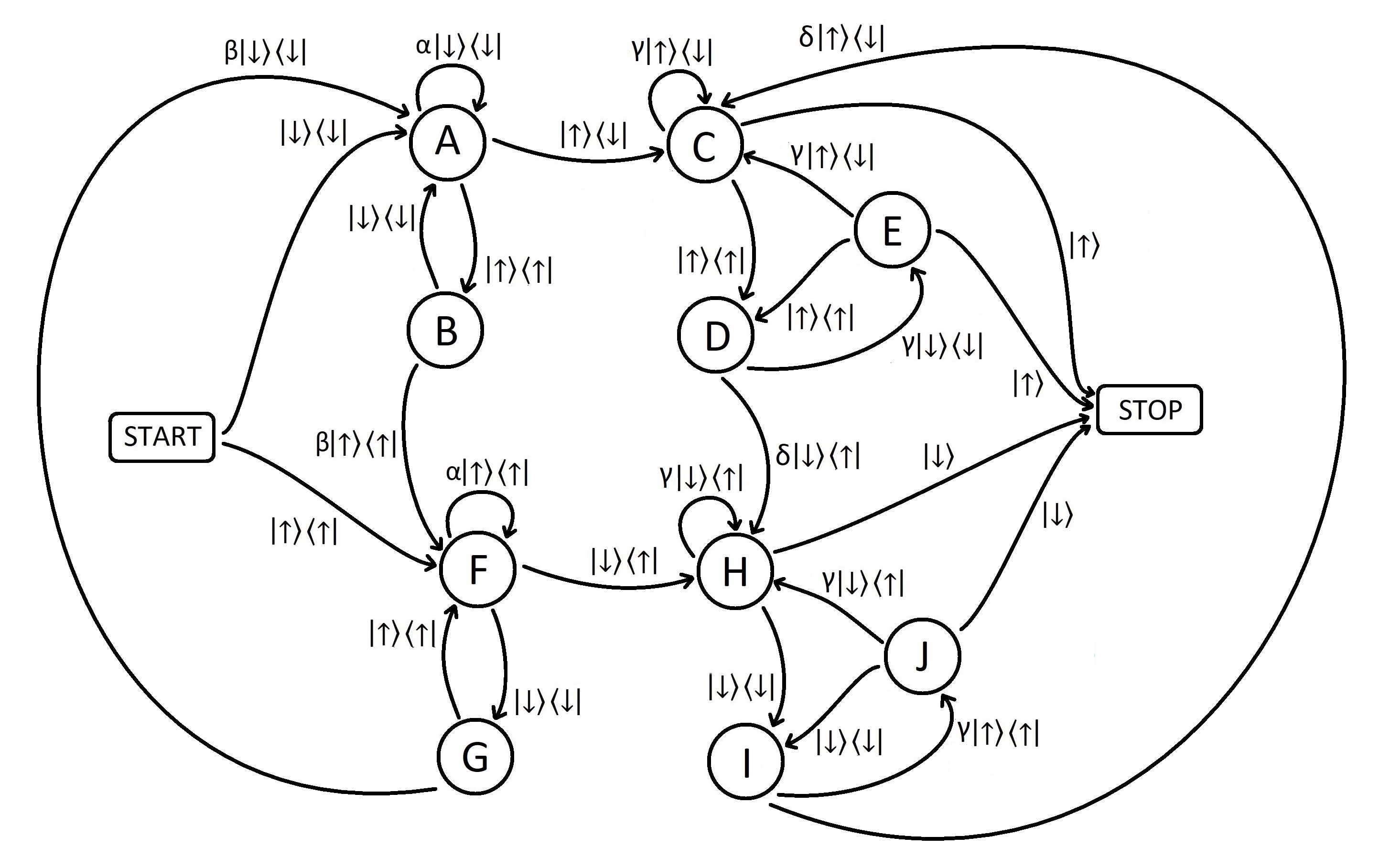}
    \caption{Automaton for the folded XXZ spin model.}
    \label{fig:fxxz-spin}
\end{figure}

Finally we construct the symmetry operator in the original model. Reproducing the spin state based on a state in the bond model is only possible if one of the spins is given due to the ambiguity of the mapping. On the level of local operators $\ket{\circ}$ and $\ket{\bullet}$ may correspond to both $\ket{\uparrow}$ or $\ket{\downarrow}$ decided only by the previous site. Consequently each element of the bond model automaton is doubled to cover both cases as presented in Figure \ref{fig:fxxz-spin}.

The two copies are located at the top and the bottom of the figure. The initial direction is decided by the first spin then the two parts are only connected through particle $\ket{2}$ identifications. The reason for this is that $\ket{0}=\ket{\bullet\bullet}$ and $\ket{1}=\ket{\circ}$ imply an even (two and zero respectively) number of spin flips whereas $\ket{2}=\ket{\bullet\circ}$ signals a single one. Therefore it changes the reference spin based on which we interpret the following one.

The elementary tensor and the boundary vectors are identified as
\begin{align}
\begin{split}
    \mathcal{L}_\text{fXXZ-spin} = & \ketbra{\downarrow}{\downarrow}
    \begin{bmatrix}
        \alpha & 1 & 0 & 0 & 0 & 0 & \beta & 0 & 0 & 0 \\
        0 & 0 & 0 & 0 & 0 & 0 & 0 & 0 & 0 & 0 \\
        0 & 0 & 0 & 0 & 0 & 0 & 0 & 0 & 0 & 0 \\
        0 & 0 & 0 & 0 & 0 & 0 & 0 & 0 & 0 & 0 \\
        0 & 0 & 0 & 1 & 0 & 0 & 0 & 0 & 0 & 0 \\
        0 & 0 & 0 & 0 & 0 & 0 & 0 & 0 & 0 & 0 \\
        0 & 0 & 0 & 0 & 0 & 1 & 0 & 0 & 0 & 0 \\
        0 & 0 & 0 & 0 & 0 & 0 & 0 & 0 & 0 & 0 \\
        0 & 0 & 0 & 0 & 0 & 0 & 0 & 1 & 0 & 1 \\
        0 & 0 & 0 & 0 & 0 & 0 & 0 & 0 & 0 & 0 \\
    \end{bmatrix}
    + \ketbra{\uparrow}{\uparrow}
    \begin{bmatrix}
        0 & 0 & 0 & 0 & 0 & 0 & 0 & 0 & 0 & 0 \\
        1 & 0 & 0 & 0 & 0 & 0 & 0 & 0 & 0 & 0 \\
        0 & 0 & 0 & 0 & 0 & 0 & 0 & 0 & 0 & 0 \\
        0 & 0 & 1 & 0 & 1 & 0 & 0 & 0 & 0 & 0 \\
        0 & 0 & 0 & 0 & 0 & 0 & 0 & 0 & 0 & 0 \\
        0 & \beta & 0 & 0 & 0 & \alpha & 1 & 0 & 0 & 0 \\
        0 & 0 & 0 & 0 & 0 & 0 & 0 & 0 & 0 & 0 \\
        0 & 0 & 0 & 0 & 0 & 0 & 0 & 0 & 0 & 0 \\
        0 & 0 & 0 & 0 & 0 & 0 & 0 & 0 & 0 & 0 \\
        0 & 0 & 0 & 0 & 0 & 0 & 0 & 0 & 1 & 0 \\
    \end{bmatrix}
    \\
    + & \ketbra{\uparrow}{\downarrow}
    \begin{bmatrix}
        0 & 0 & 0 & 0 & 0 & 0 & 0 & 0 & 0 & 0 \\
        0 & 0 & 0 & 0 & 0 & 0 & 0 & 0 & 0 & 0 \\
        1 & 0 & \gamma & 0 & \gamma & 0 & 0 & 0 & \delta & 0 \\
        0 & 0 & 0 & 0 & 0 & 0 & 0 & 0 & 0 & 0 \\
        0 & 0 & 0 & 0 & 0 & 0 & 0 & 0 & 0 & 0 \\
        0 & 0 & 0 & 0 & 0 & 0 & 0 & 0 & 0 & 0 \\
        0 & 0 & 0 & 0 & 0 & 0 & 0 & 0 & 0 & 0 \\
        0 & 0 & 0 & 0 & 0 & 0 & 0 & 0 & 0 & 0 \\
        0 & 0 & 0 & 0 & 0 & 0 & 0 & 0 & 0 & 0 \\
        0 & 0 & 0 & 0 & 0 & 0 & 0 & 0 & 0 & 0 \\
    \end{bmatrix}
    + \ketbra{\downarrow}{\uparrow}
    \begin{bmatrix}
        0 & 0 & 0 & 0 & 0 & 0 & 0 & 0 & 0 & 0 \\
        0 & 0 & 0 & 0 & 0 & 0 & 0 & 0 & 0 & 0 \\
        0 & 0 & 0 & 0 & 0 & 0 & 0 & 0 & 0 & 0 \\
        0 & 0 & 0 & 0 & 0 & 0 & 0 & 0 & 0 & 0 \\
        0 & 0 & 0 & 0 & 0 & 0 & 0 & 0 & 0 & 0 \\
        0 & 0 & 0 & 0 & 0 & 0 & 0 & 0 & 0 & 0 \\
        0 & 0 & 0 & 0 & 0 & 0 & 0 & 0 & 0 & 0 \\
        0 & 0 & 0 & \delta & 0 & 1 & 0 & \gamma & 0 & \gamma \\
        0 & 0 & 0 & 0 & 0 & 0 & 0 & 0 & 0 & 0 \\
        0 & 0 & 0 & 0 & 0 & 0 & 0 & 0 & 0 & 0 \\
    \end{bmatrix}
\end{split}
    \\
    u_\text{fXXZ-spin} = &
    \begin{bmatrix}
        \ketbra{\downarrow}{\downarrow} & 0 & 0 & 0 & 0 & \ketbra{\uparrow}{\uparrow} & 0 & 0 & 0 & 0 \\
    \end{bmatrix}
    \\
    v_\text{fXXZ-spin} = &
    \begin{bmatrix}
        0 & 0 & \ket{\uparrow} & 0 & \ket{\uparrow} & 0 & 0 & \ket{\downarrow} & 0 & \ket{\downarrow} \\
    \end{bmatrix}.
\end{align}
Note that this time the boundary vector $u$ also incorporates the action on $V_1$ in order to accommodate the reading of the first site for a reference state. Accordingly, we use the construction presented in Figure \ref{fig:open-construction3} and formula \eqref{MPO2}.

As the duplicate structure of the automaton suggests the elementary tensor can be constructed as a sum of tensor products of simpler elements. Using the identical $5\times 5$ blocks in the above expression we obtain
\begin{align}
\begin{split}
    \mathcal{L}_\text{fXXZ} = &
    \begin{bmatrix}
        \ketbra{\downarrow}{\downarrow} & 0 \\
        0 & \ketbra{\uparrow}{\uparrow} \\
    \end{bmatrix}
    \otimes
    \begin{bmatrix}
        \alpha & 1 & 0 & 0 & 0 \\
        0 & 0 & 0 & 0 & 0 \\
        0 & 0 & 0 & 0 & 0 \\
        0 & 0 & 0 & 0 & 0 \\
        0 & 0 & 0 & 1 & 0 \\
    \end{bmatrix}
    + 
    \begin{bmatrix}
        \ketbra{\uparrow}{\uparrow} & 0 \\
        0 & \ketbra{\downarrow}{\downarrow} \\
    \end{bmatrix}
    \otimes
    \begin{bmatrix}
        0 & 0 & 0 & 0 & 0 \\
        1 & 0 & 0 & 0 & 0 \\
        0 & 0 & 0 & 0 & 0 \\
        0 & 0 & 1 & 0 & 1 \\
        0 & 0 & 0 & 0 & 0 \\
    \end{bmatrix}
    \\
    + & 
    \begin{bmatrix}
        0 & \ketbra{\downarrow}{\downarrow} \\
        \ketbra{\uparrow}{\uparrow} & 0 \\
    \end{bmatrix}
    \otimes 
    \begin{bmatrix}
        0 & \beta & 0 & 0 & 0 \\
        0 & 0 & 0 & 0 & 0 \\
        0 & 0 & 0 & 0 & 0 \\
        0 & 0 & 0 & 0 & 0 \\
        0 & 0 & 0 & 0 & 0 \\
    \end{bmatrix}
    + 
    \begin{bmatrix}
        \ketbra{\uparrow}{\downarrow} & 0 \\
        0 & \ketbra{\downarrow}{\uparrow} \\
    \end{bmatrix}
    \otimes
    \begin{bmatrix}
        0 & 0 & 0 & 0 & 0 \\
        0 & 0 & 0 & 0 & 0 \\
        1 & 0 & \gamma & 0 & \gamma \\
        0 & 0 & 0 & 0 & 0 \\
        0 & 0 & 0 & 0 & 0 \\
    \end{bmatrix}
    \\
    + &
    \begin{bmatrix}
        0 & \ketbra{\uparrow}{\downarrow} \\
        \ketbra{\downarrow}{\uparrow} & 0 \\
    \end{bmatrix}
    \otimes
    \begin{bmatrix}
        0 & 0 & 0 & 0 & 0 \\
        0 & 0 & 0 & 0 & 0 \\
        0 & 0 & 0 & \delta & 0 \\
        0 & 0 & 0 & 0 & 0 \\
        0 & 0 & 0 & 0 & 0 \\
    \end{bmatrix}
\end{split}
    \\
    u_\text{fXXZ} = &
    \begin{bmatrix}
        \ketbra{\downarrow}{\downarrow} & 0 & 0 & 0 & 0 & \ketbra{\uparrow}{\uparrow} & 0 & 0 & 0 & 0 \\
    \end{bmatrix}
    \\
    v_\text{fXXZ} = &
    \begin{bmatrix}
        0 & 0 & \ket{\uparrow} & 0 & \ket{\uparrow} & 0 & 0 & \ket{\downarrow} & 0 & \ket{\downarrow} \\
    \end{bmatrix}.
\end{align}

\section{Symmetries in the 5-site Model}

\label{sec:5site}

\begin{figure}[h]
    \centering
    \includegraphics[width=12 cm]{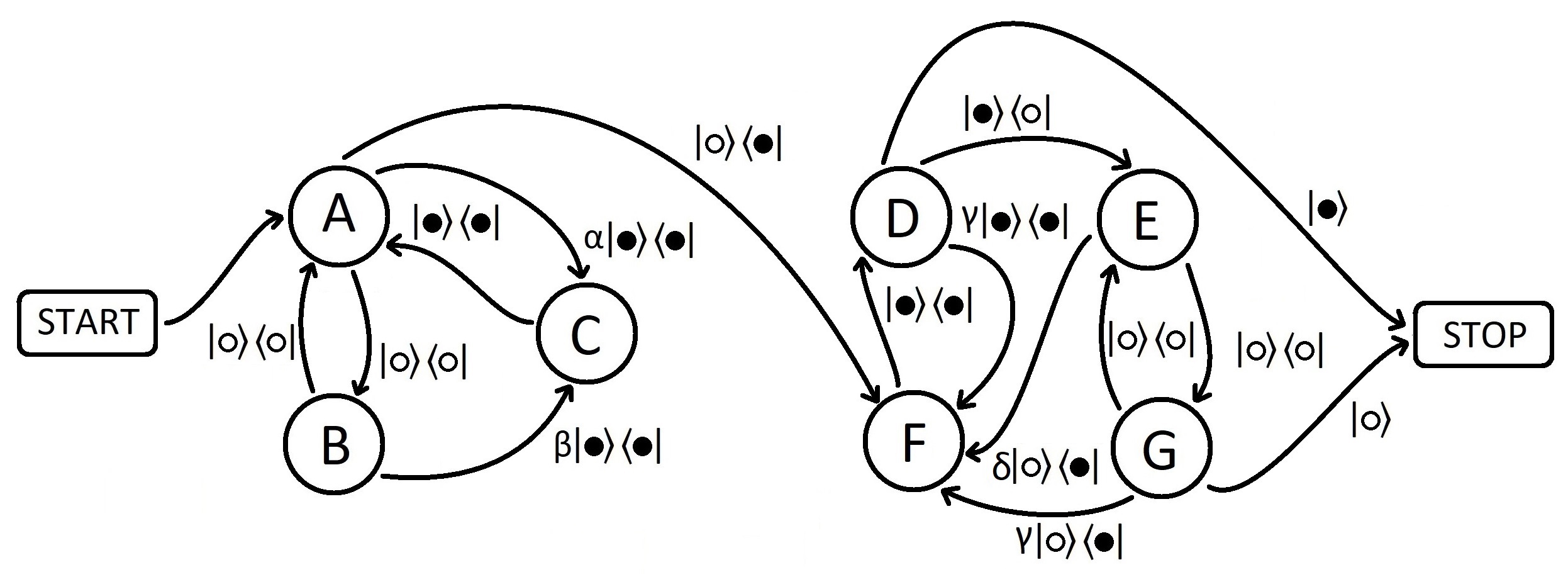}
    \caption{Automaton for the bond picture of the 5-site model.}
    \label{fig:5site-bond}
\end{figure}

The constructions is completely analogous in the other model. The automaton for the bond picture is presented in Figure \ref{fig:5site-bond}.

As the particles span one more site in the bond model, an extra automaton state is necessary both before and after the exchange that is performed through the transition from $A$ to $F$. By reading off the elements we obtain
\begin{align}
\begin{split}
    \mathcal{L}_\text{5-site}^\text{bond} = & \ketbra{\circ}{\circ} 
    \begin{bmatrix}
        0 & 1 & 0 & 0 & 0 & 0 & 0 \\
        1 & 0 & 0 & 0 & 0 & 0 & 0 \\
        0 & 0 & 0 & 0 & 0 & 0 & 0 \\
        0 & 0 & 0 & 0 & 0 & 0 & 0 \\
        0 & 0 & 0 & 0 & 0 & 0 & 1 \\
        0 & 0 & 0 & 0 & 0 & 0 & 0 \\
        0 & 0 & 0 & 0 & 1 & 0 & 0 \\
    \end{bmatrix}
    + \ketbra{\bullet}{\bullet}
    \begin{bmatrix}
        0 & 0 & 1 & 0 & 0 & 0 & 0 \\
        0 & 0 & 0 & 0 & 0 & 0 & 0 \\
        \alpha & \beta & 0 & 0 & 0 & 0 & 0 \\
        0 & 0 & 0 & 0 & 0 & 1 & 0 \\
        0 & 0 & 0 & 0 & 0 & 0 & 0 \\
        0 & 0 & 0 & \gamma & 0 & 0 & 0 \\
        0 & 0 & 0 & 0 & 0 & 0 & 0 \\
    \end{bmatrix}
    \\
    + & \ketbra{\bullet}{\circ}
    \begin{bmatrix}
        0 & 0 & 0 & 0 & 0 & 0 & 0 \\
        0 & 0 & 0 & 0 & 0 & 0 & 0 \\
        0 & 0 & 0 & 0 & 0 & 0 & 0 \\
        0 & 0 & 0 & 0 & 0 & 0 & 0 \\
        0 & 0 & 0 & 1 & 0 & 0 & 0 \\
        0 & 0 & 0 & 0 & 0 & 0 & 0 \\
        0 & 0 & 0 & 0 & 0 & 0 & 0 \\
    \end{bmatrix}
    + \ketbra{\circ}{\bullet}
    \begin{bmatrix}
        0 & 0 & 0 & 0 & 0 & 0 & 0 \\
        0 & 0 & 0 & 0 & 0 & 0 & 0 \\
        0 & 0 & 0 & 0 & 0 & 0 & 0 \\
        0 & 0 & 0 & 0 & 0 & 0 & 0 \\
        0 & 0 & 0 & 0 & 0 & 0 & 0 \\
        1 & 0 & 0 & 0 & \delta & 0 & \gamma \\
        0 & 0 & 0 & 0 & 0 & 0 & 0 \\
    \end{bmatrix}
\end{split}
    \\
    u_\text{5-site}^\text{bond} = & 
    \begin{bmatrix}
        1 & 0 & 0 & 0 & 0 & 0 & 0 \\
    \end{bmatrix}
    \\
    v_\text{5-site}^\text{bond} = &
    \begin{bmatrix}
        0 & 0 & 0 & \ket{\bullet} & 0 & 0 & \ket{\circ} \\
    \end{bmatrix}
\end{align}
As before the bond MPO in constructed according to \ref{fig:open-construction2} and formula \eqref{MPO}. States without proper MM interpretation are annihilated by the MPO again.

\begin{figure}[h]
    \centering
    \includegraphics[width=12 cm]{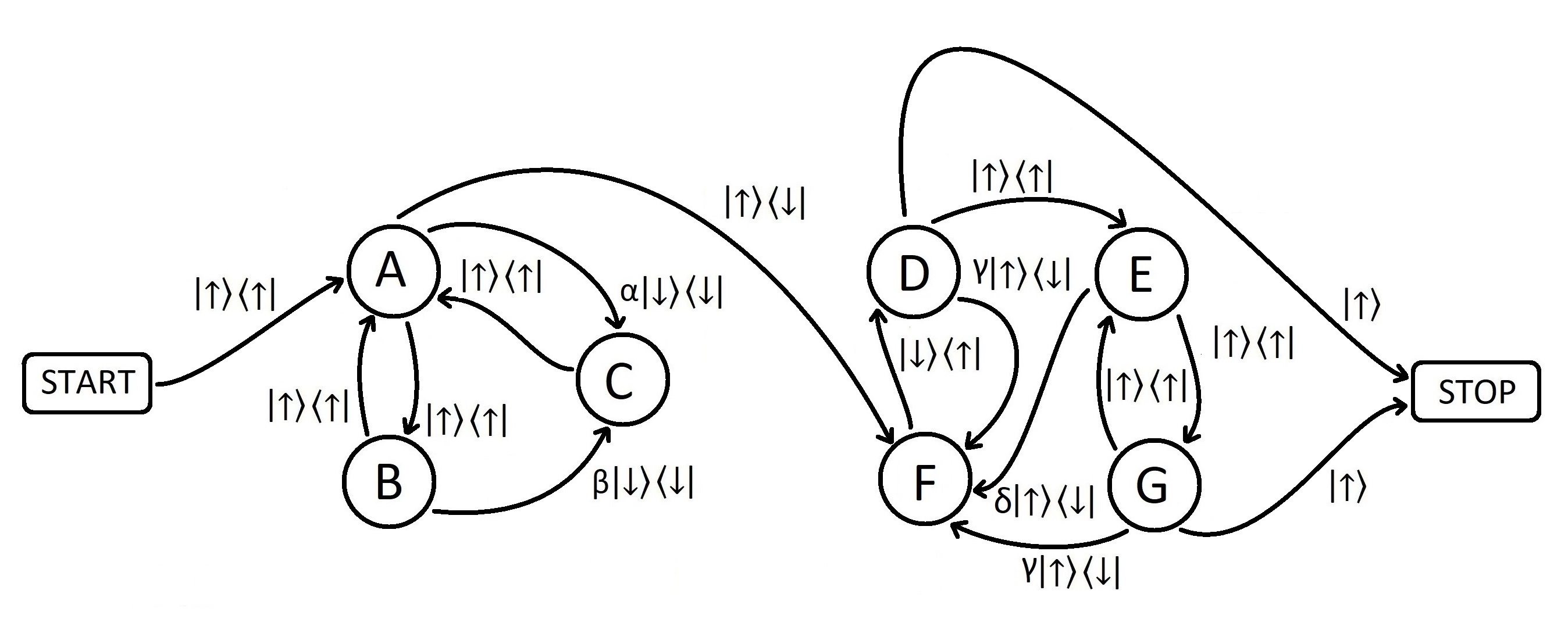}
    \caption{Automaton for the 5-site spin model.}
    \label{fig:5site-spin}
\end{figure}

Lastly we turn our attention towards the symmetry operator of the spin model. The doubling effect for the automaton does not occur here due to the unambiguous mapping between the two models ensured by the extra constraints on the Hilbert space. The state $\ket{\circ}$ always implies the sea of up spins whereas particles indicated by $\ket{\bullet\bullet}$ distinctly correspond to the isolated down spins. Consequently the automaton, shown in Figure \ref{fig:5site-spin}, has the same structure as the one for the bond model.

We remark that one could construct this automaton directly considering that whether an isolated $\ket{\downarrow}$ corresponds to $\ket{1}$ or $\ket{2}$ is decided by the parity of the number of preceding vacuum states.

In any case the elementary tensor and the boundary vectors are identified as
\begin{align}
\begin{split}
    \mathcal{L}_\text{5-site} = & \ketbra{\uparrow}{\uparrow} 
    \begin{bmatrix}
        0 & 1 & 1 & 0 & 0 & 0 & 0 \\
        1 & 0 & 0 & 0 & 0 & 0 & 0 \\
        0 & 0 & 0 & 0 & 0 & 0 & 0 \\
        0 & 0 & 0 & 0 & 0 & 0 & 0 \\
        0 & 0 & 0 & 1 & 0 & 0 & 1 \\
        0 & 0 & 0 & 0 & 0 & 0 & 0 \\
        0 & 0 & 0 & 0 & 1 & 0 & 0 \\
    \end{bmatrix}
    + \ketbra{\downarrow}{\downarrow}
    \begin{bmatrix}
        0 & 0 & 0 & 0 & 0 & 0 & 0 \\
        0 & 0 & 0 & 0 & 0 & 0 & 0 \\
        \alpha & \beta & 0 & 0 & 0 & 0 & 0 \\
        0 & 0 & 0 & 0 & 0 & 0 & 0 \\
        0 & 0 & 0 & 0 & 0 & 0 & 0 \\
        0 & 0 & 0 & 0 & 0 & 0 & 0 \\
        0 & 0 & 0 & 0 & 0 & 0 & 0 \\
    \end{bmatrix}
    \\
    + & \ketbra{\downarrow}{\uparrow}
    \begin{bmatrix}
        0 & 0 & 0 & 0 & 0 & 0 & 0 \\
        0 & 0 & 0 & 0 & 0 & 0 & 0 \\
        0 & 0 & 0 & 0 & 0 & 0 & 0 \\
        0 & 0 & 0 & 0 & 0 & 1 & 0 \\
        0 & 0 & 0 & 0 & 0 & 0 & 0 \\
        0 & 0 & 0 & 0 & 0 & 0 & 0 \\
        0 & 0 & 0 & 0 & 0 & 0 & 0 \\
    \end{bmatrix}
    + \ketbra{\uparrow}{\downarrow}
    \begin{bmatrix}
        0 & 0 & 0 & 0 & 0 & 0 & 0 \\
        0 & 0 & 0 & 0 & 0 & 0 & 0 \\
        0 & 0 & 0 & 0 & 0 & 0 & 0 \\
        0 & 0 & 0 & 0 & 0 & 0 & 0 \\
        0 & 0 & 0 & 0 & 0 & 0 & 0 \\
        1 & 0 & 0 & \gamma & \delta & 0 & \gamma \\
        0 & 0 & 0 & 0 & 0 & 0 & 0 \\
    \end{bmatrix}
\end{split}
    \\
    u_\text{5-site} = & 
    \begin{bmatrix}
        \ketbra{\uparrow}{\uparrow} & 0 & 0 & 0 & 0 & 0 & 0 \\
    \end{bmatrix}
    \\
    v_\text{5-site} = &
    \begin{bmatrix}
        0 & 0 & 0 & \ket{\uparrow} & 0 & 0 & \ket{\uparrow} \\
    \end{bmatrix}.
\end{align}
Again, $u$ has an action on $V_1$ in agreement with Figure \ref{fig:open-construction3} and equation \eqref{MPO2}.

\section{Non-integrable Perturbations}

\label{sec:nonint}

Although both  spin models that we treated are integrable it is possible to spoil integrability by adding additional terms to the Hamiltonian whilst the MPO symmetries are kept intact.

A simple choice is to use local projectors in the MM models. Adding a term constructed by the one-site operators $p=\ketbra{1}{1} + \ketbra{2}{2}$ and $n=\ketbra{0}{0}$ to the Hamiltonian density will obviously not break the symmetry since it is blind to the spin part of the wave function. We propose the following family of perturbations:
\begin{equation}
    w_j(N) = n_j n_{j+1}...\ n_{j+N-1} p_{j+N} + p_j n_{j+1} n_{j+2}...\ n_{j+N}
\end{equation}
parametrized by the integer $N\in \{ 1, 2, 3...\ \}$. This effectively counts particles that have at least $N$ preceding or following vacuum sites. Such terms do not commute with the Hamiltonian since the propagation of the particles changes the gaps between them. Additionally their implementations in the bond and spin models are relatively simple as well.

We start with the folded XXZ model. In the bond picture the corresponding perturbation is
\begin{equation}
    w_{\text{fXXZ},j}(N) = P_j^\bullet P_{j+1}^\bullet...\ P_{j+2N-1}^\bullet P_{j+2N}^\circ + P_j^\circ P_{j+1}^\bullet P_{j+2}^\bullet...\ P_{j+2N}^\bullet.
\end{equation}
This expression describes a projection to $2N$ consecutive $\ket{\bullet}$ states and a single $\ket{\circ}$ that indeed correspond to $N$ vacuum states and a single particle. The projection is blind to the particle being $\ket{1}$ or $\ket{2}$ since that is only decided by the parity of the number of $\ket{\bullet}$ states preceding the sequence according to the rules \eqref{eq:identification1}.

In the spin model we introduce the multi-site operators $\Pi_{j,k}^\pm = (1 \pm S_j^z S_k^z)/2$ that are projectors to identical and opposite spin states. Interestingly the two perturbative terms can be combined into a single one if we apply the previous projectors for non-neighboring sites:
\begin{equation}
    w_{\text{fXXZ},j}(N) = \Pi_{j,j+2N+1}^+ \prod_{k=1}^{2N-1} \Pi_{j+k,j+k+1}^-.
\end{equation}
This is interpreted as identical spins on the boundaries and alternating ones inside. The first examples are
\begin{align}
    w_{\text{fXXZ},j}(1) &= \Pi_{j,j+3}^+ \Pi_{j+1,j+2}^- \\
    w_{\text{fXXZ},j}(2) &= \Pi_{j,j+5}^+ \Pi_{j+1,j+2}^- \Pi_{j+2,j+3}^- \Pi_{j+3,j+4}^.
\end{align}
The model corresponding to $N=1$ is also integrable and known as the hard rod deformed XXZ chain. $N=2$ breaks
integrability.

Similar perturbations that keep the MPO symmetries were introduced also in \cite{maurizio-nonint-jamming}.

The 5-site model is treated analogously. According to the identification rules \eqref{eq:identification2} we have the following perturbation in the bond picture:
\begin{equation}
    w_{\text{5-site},j}(N) = P_j^\circ P_{j+1}^\circ...\ P_{j+2N-1}^\circ P_{j+2N}^\bullet P_{j+2N+1}^\bullet + P_j^\bullet P_{j+1}^\bullet P_{j+2}^\circ P_{j+3}^\circ...\ P_{j+2N+1}^\circ
\end{equation}
which has the spin model equivalent
\begin{equation}
    w_{\text{5-site},j}(N) = \Pi_{j,j+2N+2}^+ \Pi_{j+1,j+2N+1}^- \Pi_{j,j+2}^+ \prod_{k=2}^{2N-1} \Pi_{j+k,j+k+1}^+.
\end{equation}
The interpretation is that we have identical spins at the boundaries, opposite ones next to the boundaries and the rest are identical inside being also equal to the boundary ones. Again we present the first examples:
\begin{align}
    w_{\text{5-site},j}(1) =& \Pi_{j,j+5}^+ \Pi_{j+1,j+4}^- \Pi_{j,j+2}^+ \Pi_{j+2,j+3}^+ \\
    w_{\text{5-site},j}(2) =& \Pi_{j,j+6}^+ \Pi_{j+1,j+5}^- \Pi_{j,j+2}^+ \Pi_{j+2,j+3}^+ \Pi_{j+3,j+4}^+.
\end{align}

\section{Connection to Supersymmetry}

\label{sec:susy}

There is another context in which volume changing symmetries of lattice models appear, namely $\mathcal{N}=2$ supersymmetry. The corresponding Hamiltonians are generated by some supercharges $Q$ and $Q^\dag$:
\begin{equation}
    H = \{ Q, Q^\dag \},
\end{equation}
which are nilpotent: $Q^2=(Q^\dag)^2=0$. Their commutation with the Hamiltonian follows from the construction. 

Such a special structure has far reaching consequences. One can show that all energy eigenvalues satisfy $E\geq 0$,
groundstates with $E=0$ form singlet representations of the symmetry algebra whereas $E>0$ eigenstates are doublets.

One example is the well known XXZ spin chain at $\Delta=-1/2$ \cite{susy-yang-fendley}. In that case it is known that
the SUSY generators are volume changing \cite{susy-yang-fendley}. Now we show that they can be expressed as MPO's using
our techniques.

The model is defined as follows. We consider open boundary conditions and allow boundary magnetic fields so that the
Hamiltonian is given as 
\begin{equation}
    H_\text{XXZ} = - h (S_1^z + S_L^z) - \sum_{j=1}^{L-1} \Big( S_j^x S_{j+1}^x + S_j^y S_{j+1}^y + \Delta S_j^z S_{j+1}^z \Big).
\end{equation}
The corresponding supercharge
\begin{equation}
    Q = \sum_{j=1}^L Q_j
\end{equation}
exists only in the form of a volume changing operator. We present the original construction then use our framework to produce it as an MPO as well.

Due to the fermionic nature of the symmetry operators it is worth using a fermionic representation of the spins. We introduce two species of fermions $f_{j,\uparrow}$ and $f_{j,\downarrow}$ and demand that each site must be occupied by exactly one of them. The spin operators are represented as
\begin{equation}
    \underline{S}_j = f_{j,\mu}^\dag \underline{\sigma}_{\mu,\nu} f_{j,\nu},
\end{equation}
where $\underline{\sigma}$ stands for the Pauli matrices and $\mu, \nu \in \{ \downarrow, \uparrow \}$. The raising and lowering operators consequently become
\begin{equation}
    S_j^+ = f_{j,\uparrow}^\dag f_{j,\downarrow}, \qquad S_j^- = f_{j,\downarrow}^\dag f_{j,\uparrow}.
\end{equation}
Finally we introduce the shift operators $A_j^{R\dag}$ which move all the fermions on sites $k>j$ to the right by one. It thus creates an extra site on the right and leaves site $j+1$ unoccupied. We are now ready to present the supercharge density:
\begin{equation}
    Q_j = f_{j,\uparrow}^\dag f_{j+1,\uparrow}^\dag f_{j,\downarrow} A_j^{R\dag},
\end{equation}
which practically performs the exchange $\ket{\downarrow} \rightarrow \ket{\uparrow\uparrow}$ at site $j$. Regarding the fermionic sign $Q_j$ introduces a factor of $(-1)^{j-1}$. Note that with $Q$ adding an extra site and $Q^\dag$ taking one away the Hamiltonian indeed preserves the system size.

\begin{figure}[h]
    \centering
    \includegraphics[width=12 cm]{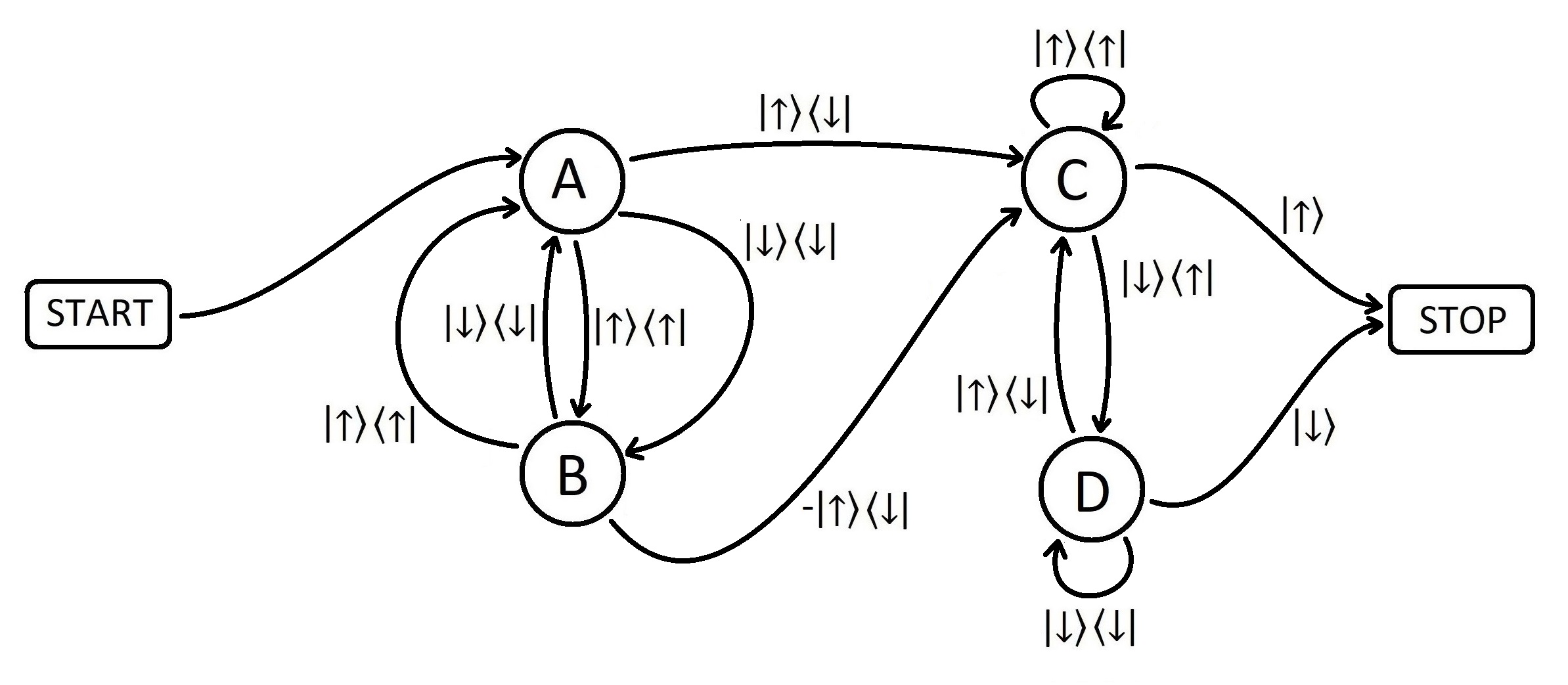}
    \caption{Automaton for XXZ supercharge.}
    \label{fig:susy}
\end{figure}

To construct the supersymmetry operator within our framework, we need an automaton that first acts identically keeping a record of the parity of the number of sites then performs the exchange $\ket{\downarrow} \rightarrow \ket{\uparrow\uparrow}$ and adds the necessary sign. After that it shifts the spins to the right by one and finishes with adding the extra site. This is shown in Figure \ref{fig:susy}.

Automaton state $A/B$ corresponds to an even/odd number of sites located left to the modified one. The elementary tensor and the boundary vectors are easily read off:
\begin{align}
\begin{split}
    \mathcal{L}_\text{SUSY} = & \ketbra{\uparrow}{\uparrow}
    \begin{bmatrix}
        0 & 1 & 0 & 0 \\
        1 & 0 & 0 & 0 \\
        0 & 0 & 1 & 0 \\
        0 & 0 & 0 & 0 \\
    \end{bmatrix}
    + \ketbra{\downarrow}{\downarrow}
    \begin{bmatrix}
        0 & 1 & 0 & 0 \\
        1 & 0 & 0 & 0 \\
        0 & 0 & 0 & 0 \\
        0 & 0 & 0 & 1 \\
    \end{bmatrix}
    \\
    + & \ketbra{\uparrow}{\downarrow}
    \begin{bmatrix}
        0 & 0 & 1 & 0 \\
        0 & 0 & -1 & 0 \\
        0 & 0 & 0 & 0 \\
        0 & 0 & 1 & 0 \\
    \end{bmatrix}
    +\ketbra{\downarrow}{\uparrow}
    \begin{bmatrix}
        0 & 0 & 0 & 0 \\
        0 & 0 & 0 & 0 \\
        0 & 0 & 0 & 1 \\
        0 & 0 & 0 & 0 \\
    \end{bmatrix}
\end{split}
    \\
    u_\text{SUSY} = & 
    \begin{bmatrix}
        1 & 0 & 0 & 0 \\
    \end{bmatrix}
    \\
    v_\text{SUSY} = & 
    \begin{bmatrix}
        0 & 0 & \ket{\uparrow} & \ket{\downarrow} \\
    \end{bmatrix}.
\end{align}
Once again $v \in V_a \otimes V_{L+1}$ in order to add the extra site. Accordingly the construction presented in equation \eqref{MPO} and Figure \ref{fig:open-construction2} is used.

\section{Conclusions}

\label{sec:concl}

In this work we discussed three different classes of volume changing symmetries described by generalized Matrix Product
Operators. The first two classes belong to the folded XXZ model (and its perturbations), and to a new hopping model. The
third class consists of the known supersymmetry of the XXZ model at $\Delta=-1/2$; our contribution here is to express
that known symmetry as a generalized MPO.

One of the most interesting questions is to classify the various mechanisms that can lead to volume changing
symmetries. The examples that we presented here were found on a one by one basis.
All of them involve a local operation which exchanges local segments in the chain, accompanied by a shift of the
remainder of the chain. If such operations are meant to commute with a local Hamiltonian, then clearly an extreme fine
tuning is required. However, our examples show that even such exotic symmetries allow for multiple types of
realizations, beyond those given by supersymmetry. It would be desirable to develop a systematic treatment for this problem.

The symmetry operators that we constructed bear some similarity with the constructions in the works
\cite{fendley-relations,verstraete-dualities-1}: the common point is that our symmetry operators are locally constructed
in the form of a tensor network, and strictly speaking they act between two different Hilbert spaces, just like the
duality transformations treated in those works. It would be interesting to find more direct connections between our
results and the works \cite{fendley-relations,verstraete-dualities-1}.

\section*{Acknowledgements}

This work was supported by the
Hungarian National Research, Development and Innovation Office, NKFIH Grant No. K-145904 and the 
NKFIH excellence grant TKP2021-NKTA-64.


\providecommand{\href}[2]{#2}\begingroup\raggedright\endgroup

\end{document}